\documentclass[10pt]{article}
\usepackage{graphicx}
\usepackage{amsmath}
\usepackage{amsfonts}
\usepackage{arxiv}

\begin{document}
\title{True 3D reconstruction in digital holography}

\author{
  Jasleen Birdi, Sunaina, Mansi Butola and Kedar Khare* \\
  Department of Physics, Indian Institute of Technology Delhi, New Delhi 110016 India \\
  \texttt{*kedark@physics.iitd.ac.in}\\
 }

\maketitle

\begin{abstract}
We examine the nature of the 3D image as obtained by replay (or back-propagation) of the object wave from the hologram recording plane to the original object volume. While recording of a hologram involves transferring information from a 3D volume to a 2D detector plane, the replay of the hologram involves creating information in a set of 3D voxels from a much smaller number of 2D detector pixels, which at first look appears to be surprising. We point out that the hologram replay process is a Hermitian transpose (and not inverse) of the hologram formation process and therefore only provides an approximation to the original 3D object function. With the knowledge of this Hermitian transpose property, we show how one may realize true 3D image reconstruction via a regularized optimization algorithm. The  numerical illustrations of this optimization approach as presented here show excellent slice-by-slice tomographic 3D reconstruction of the original object under the weak scattering approximation. In particular, the reconstructed 3D image field has near-zero numerical values at voxels where the original object did not exist. We note that 3D image reconstruction of this kind cannot be achieved by the traditional physical replay process. In this sense the proposed methodology for digital holographic image reconstruction goes beyond numerically mimicking the traditional film based holographic replay. The reconstruction approach may find potential applications in a number of digital holographic imaging systems. 
\end{abstract}

% keywords can be removed
\keywords{Holography, 3D imaging, computational imaging, image reconstruction}

\section{Introduction}
Dennis Gabor's work \cite{gabor1948new} on the improvement of the electron microscope in $1947$ led to the remarkable invention of the principle of holographic imaging. Since the hologram is an interference record, it captures information about amplitude as well as phase of the unknown object wave unlike the conventional photography which records intensity-only information. The replay field of the film based hologram has one component that corresponds to the back-propagating object wave. This component focuses back at the location of the original object leading to image formation. The visual inspection of this replay field gives one a perception of the reconstruction of the original three-dimensional (3D) object as our eyes are very good at concentrating on the focused image while ignoring any diffuse background. This 3D imaging aspect has always been an attractive feature of holography and is often highlighted in textbooks as well as classrooms \cite{hecht2002optics}. 

For the last twenty five years, film based holography has slowly been replaced by digital holography \cite{schnars2015digital, kim2006interference, kim2010} where the hologram/intereference pattern is recorded on a digital sensor (e.g. CCD or CMOS array) and the image reconstruction task is performed numerically. One of the advantages of digital holography is that the interference data is now available in numerical form. Once the complex-valued object wave in the hologram plane is recovered from the interference data, the 3D image reconstruction is usually performed by mimicking the film based hologram replay that consists of propagating the object wave field at the hologram plane back to the location of the original object volume. While the replay in film based holography was primarily used for applications involving visual observation of images, digital holography offers the possibility of quantitative imaging. This aspect has particularly become important in applications of digital holography to microscopy where the phase images of biological cells are used for developing diagnostic tools \cite{Park2018, mangal2019}. While a lot of literature has been devoted to digital holography and its various applications over the past few decades, the nature of the 3D reconstruction offered by numerical back-propagation of the object wave still needs careful examination in our opinion. The terminology ``3D imaging'' is in fact often used somewhat incorrectly. For example, the typical phase images of cells obtained using a digital holographic microscope are often rendered as surface plots. The nature of 3D image representation in such cases is however not similar to a z-stack of slices as may be obtained from a confocal microscope. 

One may begin the probing of this topic by noting that the holographic recording involves storing information about objects in a 3D volume onto a 2D digital array detector in the form of an interference pattern. A number of algorithmic methodologies have evolved over the past few decades that allow the recovery of the object field in the recorded hologram plane. For example, the off-axis holograms are typically processed using the Fourier transform method where the object beam information is recovered by filtering the region near the cross-term peak in 2D Fourier transform of the hologram data frame \cite{takeda1982fourier}. The on-axis digital holograms on the other hand are typically processed using the multi-shot phase shifting framework \cite{creath1988phase, yamaguchi1997phase}. More recently, optimization methodologies have been developed that allow accurate full resolution object wave recovery from a single hologram frame uniformly for both on-axis as well as off-axis geometries \cite{singh2015PRA, singh2017single, rajora2019mean}. The holographic replay of the object wave is a linear operation on the object wave field. If this operation truly reconstructed the original 3D object, it would mean recovery of voxels in the original object volume using comparatively a much smaller number of pixels in the hologram plane, which appears to be problematic. It is actually known from early days of holography that the back-propagated field is only an approximation to the original 3D object. In Gabor's own description \cite{gabor1971holography}, the term three-dimensional was only used ``to designate an image (or the process which produced it) that is substantially isomorphic with the original object''. In a film-based holography setup there is not much one can do beyond physical replay of the field for image reconstruction, however, this need not be the case when object field in the hologram plane is available numerically. It is therefore important to investigate if a true 3D object reconstruction better than the usual back-propagation is even possible using the 2D object field information at the hologram plane. The question we investigate here is relevant to a number of allied problems such as holographic memory involving data storage in multiple layers \cite{hong1995volume, sheridan_SciRep}, Fresnel incoherent holographic imaging \cite{rosen2007digital}, integral imaging \cite{park2009recent, xiao2013advances}, lightfield imaging \cite{ihrke2016principles}, and optical metrology \cite{malacara2007optical} where 3D information is typically needed using data recorded on a 2D detector array.  

The problem of 3D image reconstruction has received attention particularly in case of in-line particulate holograms. By simple back-propagation of a particle hologram, it is possible to observe locations in space where the particles come into focus and localize them \cite{Kreis1997}. A more detailed approach involves fitting of Mie scattering models to particulate holograms \cite{fung2012imaging} and this method has been shown to localize particulates in colloids. The problem of 3D field reconstruction has also been cast in the inverse problems framework mainly for in-line hologram data associated with particulates and thin thread-like objects \cite{soulez2007inverse, brady2009compressive, denis2009inline, mallery2019regularized}. In these approaches, the data is the recorded interference (intensity) pattern and the associated optimization problem aims to find a sharp focused object that agrees with the recorded intensity data with image domain constraints such as  reality, positivity and sparsity. In these works, the non-linear term corresponding to object wave energy is assumed to be small (and treated as model error), and the data contains linear terms corresponding to both the object wave and its conjugate. The back-propagation of this data vector therefore contains both the focused object wave and its diverging twin. The role of constraints in these works is mainly to enforce object sparsity, however, implicitly the sparsity is also expected to eliminate the twin image. A de-convolution approach for 3D reconstruction has been presented in \cite{latychevskaia2010depth}. More recently, a deep learning based method has been developed for localization of particulates from their in-line holograms and has shown promising results \cite{shimobaba}. An iterative method using two in-line intensity measurements for 3D reconstruction has also been demonstrated recently \cite{latychevskaia2019}. There is considerable interest in 3D slice-by-slice phase reconstruction of volume objects like biological cells. As per the fundamental theorem of diffraction tomography \cite{Wolf1969}, knowledge of scattered field on two planes on either side of a weakly scattering object is sufficient to provide a low-pass filtered version of its 3D structure. This problem has also been handled using phase tomography approach where phase projections of a transparent 3D object are obtained by recording holograms of the same object from a number of view angles as allowed by tilt or other mechanisms \cite{Cheng2010,choi2007tomographic}. The phase projections are then used for tomographic image reconstruction.  
 
Our aim in this work is to formulate the 3D reconstruction as an inverse problem using the numerically recovered object wave in the hologram plane as the data. First of all this approach will not require us to worry about the twin image elimination issues \cite{zhang2018twin}, and as a result the nature of back-propagated object field in the original object volume can be understood in a simpler manner. Secondly, we do not make any assumption that the energy in non-linear term in the hologram is small. The treatment presented here is therefore independent of the hologram recording configuration (in-line or off-line). We find that this framework provides a simple analytical result that the back-propagation of the object field from hologram plane is the Hermitian transpose of the hologram formation process. Further, this association is very useful in implementation of an iterative 3D image reconstruction algorithm that goes back and forth between the 3D object domain and the 2D detector domain. The resultant 3D image reconstruction is also observed to be a significantly better quantitative estimate of the original 3D object compared to what is obtained by the traditional holographic replay process.                  

This paper is organized as follows. In section $2$, we describe the true $3$D reconstruction problem, firstly, in an intuitive way that is followed by a mathematical analysis of the problem. Section $3$ provides the methodological description of the approach adopted by us. Our numerical experimentation results for amplitude and phase test objects are shown in section $4$. Finally, in section $5$, we provide concluding remarks.

\section{Problem overview}
\subsection{Physical intuition}
For illustrative purpose let us consider a 3D object volume consisting of multiple planes with four point objects (see Fig. 1(a)). A spatially coherent monochromatic source is split into two parts. One part reflects off the four point objects generating an object wave and the other part acts as a tilted plane reference beam. At the 2D sensor plane, the interference pattern due to the total field from the four sources and the plane reference beam is recorded. We make an assumption that only the primary scattered field from each of the four objects predominantly forms the total object wave at the sensor. This assumption is reasonable and has been used since the early days of holography \cite{gabor1956theory}. In the present work, we assume that the object field at the  sensor/hologram plane is known using single-shot or multi-shot methods for complex object wave recovery in digital holography \cite{yamaguchi1997phase, singh2015PRA, rajora2019mean}. The task is to reconstruct the 3D object distribution from the 2D object field at the hologram plane. The conventional holographic techniques rely on illuminating the hologram with the conjugate reference beam. This leads to back-propagation of the object wave to the different planes of 3D volume as shown in Fig. 1(b). For illustration, we have shown the back-propagated field amplitudes corresponding to different particles in the original 3D volume with different colors. It is clear that in any plane where one of the point objects gets focused, there are de-focused versions of fields due to the other objects as well. This physical hologram reconstruction process can be mimicked for 3D image reconstruction in digital holography by numerical back-propagation of the object field. It is then evident that while the four point objects exist at specific locations within the volume as in Fig. 1(a), the back-propagated field is non-zero everywhere. The operation of back-propagation of the object wave from the hologram plane therefore does not provide a true 3D distribution of the original object. The nature of back-propagated object wave will be explained in more detail in the following discussion.          
\begin{figure}[htbp]
\centering
\includegraphics[width=0.9\textwidth]{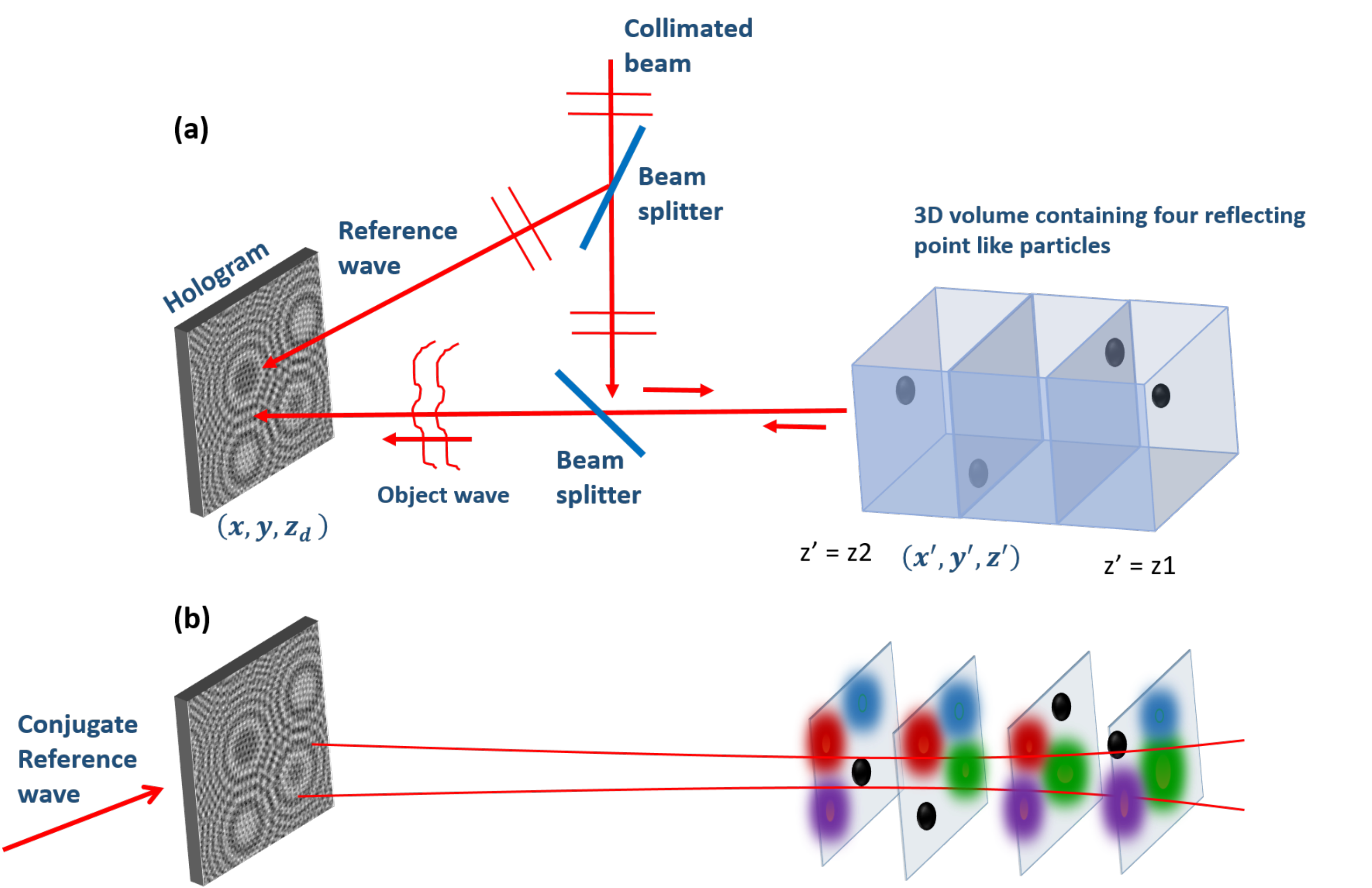}
\caption{Pictorial representation of hologram recording and replay processes. (a) Hologram recording of a 3D object, consisting of $4$ reflective point objects located at different planes, with a tilted plane reference wave. (b) Back-propagation of the known 2D object field at the hologram plane to the different planes of the 3D volume. In each plane, one can clearly see the focused as well as the de-focused field distribution. In (b), colors are used for the purpose of representing the de-focusing effect.}
\end{figure}
\subsection{Mathematical Analysis}
\begin{figure}[htbp]
    \centering
    \includegraphics[width=0.9\textwidth]{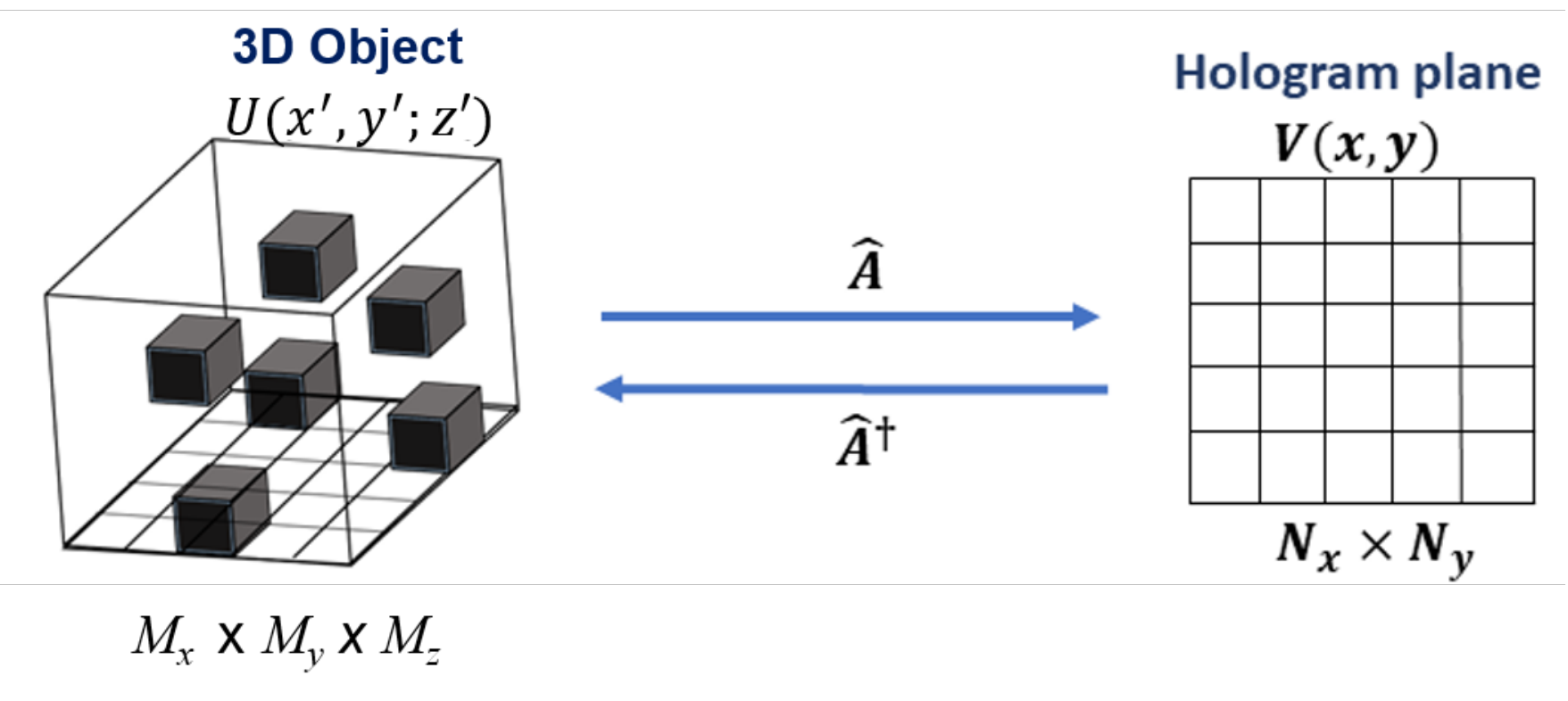}
        \caption{Pictorial representation of hologram recording and back-propagation of the object field to the original object volume. The operator $\hat{A}$ denotes the hologram formation process which takes information about a 3D volume to a 2D detector plane. The back-propagation of the object field is seen to be equivalent to the adjoint or $\hat{A}^{\dagger}$ operation.}
        \label{fig:AAdagger}
\end{figure}

Let the $3$D complex-valued object function be denoted by $U(x',y'; z')$. Numerical values of $U(x',y'; z')$ represent the reflection (or the transmission) coefficient of the voxel at $(x', y', z')$ depending on the hologram recording geometry. The voxels in $U(x',y'; z')$ where no object is present are equal to zero. In particular, if the object is a reflective surface represented as $z' = f(x',y')$, then the object function will be represented as:
\begin{equation}
    U(x',y';z') = r(x',y') \;\; \delta[z' - f(x',y')],
\end{equation}
where $r(x',y')$ denotes the reflection coefficient of the object at location $(x',y', z' = f(x',y'))$. The forward propagation of the object field to the hologram plane can be represented by a linear operator $\hat{A}$:
\begin{equation}\label{eq:A}
 V (x,y)  = \hat{A} \, U (x^\prime, y^\prime; z^\prime).
\end{equation}
In accordance with \cite{gabor1956theory} we neglect the secondary scattering, and the operator $\hat{A}$ thus consists of propagating each slice in the function $U(x',y'; z')$ to the detector plane (located at $z = z_d$) and adding the total field to get a complex valued function denoted by $V(x,y)$ as follows:
\begin{equation}\label{eq:conv}
V(x,y)  = \int_{z_1}^{z_2} dz' e^{ik(z'-z_2)} \int\int dx' dy' U(x',y'; z') h(x-x',y-y';z_d - z').
 \end{equation}
 The factor $\exp[ik(z'-z_2)]$ arises in the reflection geometry shown in Fig. 1(a) as the illuminating beam has different relative phases on different object planes. The limits $z_1$ and $z_2$ on integration over $z'$ denote the $z$-range of the volume of interest. These limits will be omitted in the following discussion for brevity. The impulse response $h(x,y;z)$ in a paraxial geometry may be assumed to be that for Fresnel diffraction \cite{goodman2005introduction}:
 \begin{equation}
     h(x,y;z) = \frac{1}{i\lambda z} \exp(ikz) \exp[i\frac{\pi}{\lambda z} (x^2 +y^2)].
 \end{equation}
Here $\lambda$ is the wavelength of illumination and $k = 2\pi/\lambda$ is the wave-number. For a more general wide-angle case, the preferred impulse response may be obtained via the angular spectrum approach.
Next we proceed to determine the adjoint operation $\hat{A}^{\dagger}$ using the scalar product relation:
\begin{equation}\label{eq:2d3dsp}
    <V_1, \hat{A}\, U_2>_{2D} \, = \, <\hat{A}^\dagger V_1, U_2>_{3D}.
\end{equation}
The suffixes 2D and 3D on the scalar products tell us that the scalar product on the LHS is evaluated on the detector coordinates whereas the scalar product on the RHS is evaluated over the 3D volume of interest. In the above equation the detector field $V_1 = \hat{A} U_1$ and $U_1, U_2$ are any two 3D object functions. Solving the LHS of the above equation and using Eq.(\ref{eq:conv}), we have
\begin{align}\label{eq:innerp}
&< V_1, \hat{A}\: U_2>  \nonumber \\ & = \int \int dxdy  V_1^*(x,y)\int dz' e^{ik(z'-z_2)} \int\int dx' dy'  U_2 (x', y';z') h(x-x',y-y';z_d - z').
 \end{align}
Rearranging the order of integration we can write the above relation as:
\begin{align}\label{eq:innerp1}
&< V_1, \hat{A}\: U_2>  \nonumber \\ & = \int \int\int dz' dx' dy' [\int\int dx dy V_1 (x,y) e^{-ik(z'-z_2)} h^{\ast}(x-x',y-y';z_d - z')]^{\ast} U_2 (x', y';z').
\end{align}
Using Eqs. (\ref{eq:2d3dsp}) and (\ref{eq:innerp1}), we observe that the adjoint operator $\hat{A}^{\dagger}$ acting on the detector field $V_1(x,y)$ may be defined as:
\begin{equation}\label{eq:Adagger}
[\hat{A}^{\dagger} V_1] (x',y';z')  = e^{-ik(z'-z_2)} \int\int dx dy \; V_1 (x,y)  h^{\ast}(x-x',y-y';z_d - z').
\end{equation}
Apart from the phase factor $e^{-ik(z'-z_2)}$ which nullifies the initial phase on different object planes due to illumination, we observe that the $\hat{A}^{\dagger}$ operation simply involves the back-propagation (as denoted by $h^{\ast}$) from detector plane to all the planes $z = z'$ in the volume of interest. The operations corresponding to the hologram formation and the back-propagation of the object field are shown pictorially in Fig. \ref{fig:AAdagger} for clarity. 
The traditional holographic replay of the object field is thus equivalent to the Hermitian transpose of the forward hologram formation operation denoted by operator $\hat{A}$. In fact since operator $\hat{A}$ takes voxels in 3D object space to a much smaller number of pixels in 2D, the formal inverse operation $\hat{A}^{-1}$ does not exist. Obtaining a true 3D solution which is non-zero only at the voxels where the original object existed may however be possible using a regularized optimization framework where appropriate constraints representing the desirable properties of the object function are modelled. We proceed to do this in the next section. It is important to note that such true 3D image reconstruction cannot be generated by any physical replay. Applying diffraction theory to the back-propagating field we can clearly observe that if the replay field was identically zero in some plane (where the original object did not exist), then it has to be zero on every subsequent plane. In this sense the optimization based solution we construct in the following section takes us away from the traditional picture of physical hologram replay.   
\section{True 3D recovery: Methodology}
From the previous section, firstly, it is now clear that the back-propagated object wave constitutes the $\hat{A}^\dagger$ operation. Secondly, due to the nature of the reconstruction problem where number of detector pixels is typically much smaller compared to the number of voxels in the 3D volume of interest, it may not be possible to obtain a true 3D solution in the most general case. The solution may however be feasible in case the object function has sufficient sparsity in the voxel domain or some transform domain. Such constraints if applicable to the object under consideration will allow us to reconstruct the true solution. For example, if the object consists of a collection of point sources as in case of particulate holograms, $\ell_1$ norm penalty may be suitable for imposition of solution sparsity. For a variable $\overline{U}$, it is defined as
\begin{equation}
    \| \overline{U} \|_1 = \sum_{a=1}^{M_x} \sum_{b=1}^{M_y} \sum_{c=1}^{M_z} |\overline{U}_{a,b,c}|,
\end{equation}
where indices $(a,b,c)$ run over $x,y$ and $z-$axes, respectively, and $M = M_x \times M_y \times M_z$ denotes the total number of voxels in the 3D object.
On the other hand, if the object is sparse in the image gradient domain, its sparsity can be enforced by the total variation (TV) penalty \cite{rudin1992nonlinear}. For simplicity, we consider a TV penalty for each of the constituting 2D planes, which for every $c^{th}$ plane, denoted by $\overline{U}_{:,:,c}$, is given by
\begin{equation}
     \| \overline{U}_{:,:,c} \|_{TV} = \sum_{a=1}^{M_x} \sum_{b=1}^{M_y} \sqrt{ \big| \overline{U}_{a,b,c} - 
      \overline{U}_{a-1,b,c} \big|^2 + \big| \overline{U}_{a,b,c} -  \overline{U}_{a,b-1,c} \big|^2}.
\end{equation}
More general forms of gradient based penalties such as Huber penalty may also be applicable. Depending upon the object of interest, the gradient based penalty may be applied in all three directions. In addition to the sparsity constraint, the operation $\hat{A}$ applied to the object $\overline{U}$ is expected to be consistent with the given object field at the hologram plane. Keeping this in mind, we propose to solve the following minimization problem to get an estimate $\overline{U}$ of the 3D field:
\begin{equation} \label{eq:cost_fun}
    \underset{\overline{U}}{\rm{minimize}} \,\, C_1 (\overline{U},{\overline{U}}^{\;\ast}) + \alpha \, C_2 (\overline{U}, {\overline{U}}^{\;\ast}).
\end{equation}
The first term $C_1$ is given by 
\begin{equation}
C_1 (\overline{U},{\overline{U}}^{\;\ast}) = \frac{1}{2}{\| V - \hat{A} \,\overline{U} \|_F^2}    
\end{equation}
 that ensures the data consistency of the estimated 3D field. Here $V$ is the data representing the 2D object field at the hologram plane (as may be derived from the recorded hologram), $\hat{A}$ is the operator defined in Eq. (\ref{eq:conv}), and $\|.\|_F$ denotes the Frobenius norm of its argument. The second term $C_2$ in the above problem is the chosen sparsity regularization term. The real valued quantity $\alpha > 0$ is the regularization parameter that determines the relative weight between the two terms of the cost function. 

In problem (\ref{eq:cost_fun}), the term $C_1$ is Lipschitz-differentiable, whereas $C_2$, whether chosen to be $\ell_1$ norm or TV penalty, is non-differentiable. In the optimization literature, this type of problem has recently been dealt with in an iterative manner. In each iteration, a gradient descent step is applied on $C_1$, followed by a proximity step on $C_2$, which is referred to as the forward-backward structure. Based on this, we solve problem (\ref{eq:cost_fun}) using an accelerated optimization algorithm encompassing forward-backward like iterations, namely fast iterative shrinkage-thresholding algorithm (FISTA) \cite{beck2009fast}. The main steps involved in this algorithm are described in the following.
% Gradient step
We initialize $\overline{U}$ with the guess solution $\overline{U}^{(0)}$ and consider $Z^{(1)}  = \overline{U}^{(0)}$. At $n^{\rm{th}}$ iteration, gradient descent scheme consists of updating the solution by going in the descent direction obtained by the gradient of the first term $C_1$ of the cost function to be minimized, i.e.
\begin{equation} \label{eq:grad}
    \widetilde{U}^{(n)}= Z^{(n)}- \tau \, [\nabla C_1]_{Z^{(n)}},
\end{equation}
where $- [\nabla C_1]_{Z^{(n)}} =  - \hat{A}^{\dagger} (\hat{A} \, Z^{(n)} - V)$ corresponds to the gradient descent term evaluated at $Z^{(n)}$. The step size $\tau$ is chosen as $\tau = 1/\kappa $, where $\kappa = \| \hat{A}^{\dagger} \hat{A} \|_S$ with $\|.\|_S$ denoting the spectral norm of its argument. With this choice, $\kappa$ is essentially the Lipschitz constant associated with the Lipschitz-differentiable function $C_1$. The spectral norm $\kappa$ may be evaluated using the power iteration method. Once the gradient-updated solution $\widetilde{U}^{(n)}$ is available, the proximity operator of the second term in problem (\ref{eq:cost_fun}) needs to be evaluated. By definition, proximity operator of $\alpha \, C_2$ at $\widetilde{U}^{(n)}$ is given by
\begin{equation} \label{eq:prox_def}
    {\rm prox}_{\alpha \, C_2}(\widetilde{U}^{(n)}) = \underset{L}{\rm{argmin}} \, \, \alpha \, C_2 (L) + \frac{1}{2} \| L - \widetilde{U}^{(n)} \|_F^2.
\end{equation}
Initially introduced by Moreau \cite{moreau1965proximite}, this operator is widely used to tackle non-smooth functions in a variety of signal and image processing problems. To get an idea behind this operator, note that the quadratic term in Eq. (\ref{eq:prox_def}) restricts the solution from going too far from $\widetilde{U}^{(n)}$ while minimizing for $C_2$ as well. Thus, the gradient step together with the proximity step aims to find the minimizer of problem of the form (\ref{eq:cost_fun}). As previously discussed, the chosen prior for a sparse object consisting of point sources is the $\ell_1$ norm penalty. In addition, if the object is an amplitude object, the reality and positivity constraints may be imposed. This can be enforced by the use of indicator function $1_{\mathbb{R}_+^{N}}(.)$ of real and positive orthant. Thus, the regularization term in this case is given by: $C_2 (.) =  \| . \|_1 + 1_{\mathbb{R}_+^{N}} (.)$. The associated proximity operator reduces to the positive soft-thresholding operator, $\Gamma$ \cite{combettes2011proximal}. In particular, we set $\Gamma_{\mu} (\widetilde{U}^{(n)}) = \overline{U}^{(n+1)}$ such that for all $(a,b,c) \in (M_x,M_y,M_z)$,
\begin{equation}
 [\overline{U}^{(n+1)}]_{a,b,c}  = \begin{cases}
& \text{Re}\big([\widetilde{U}^{(n)}]_{a,b,c} \big) - \mu, \quad \text{if} \quad \text{Re} \big( [\widetilde{U}^{(n)}]_{a,b,c} \big) \geq \mu \\
&  0, \qquad \qquad \qquad \qquad  \text{otherwise},
\end{cases}
\end{equation}
where $\rm{Re} (.)$ denotes the real part operator and $\mu = \alpha \tau$ in the considered case. Intuitively, the positive soft-thresholding operator sets all values less than the thresholding parameter $\mu$ to zero, while shrinking the rest of the values by an amount $\mu$. As a result, iteration-by-iteration, it leads to promoting sparsity and positivity of its argument. For the second considered case, where $C_2(.) = (.)_{TV}$, the proximity operator does not have a closed form solution. In fact, it needs to be computed using sub-iterations, for which we use the fast gradient
projection (FGP) method proposed in \cite{beck2009fasttv}.

Now, once we get an updated $\overline{U}^{(n+1)}$, the peculiarity of FISTA is that in every iteration, the gradient scheme is employed on a linear combination of last two updated values $\overline{U}^{(n+1)}$ and $\overline{U}^{(n)}$, for faster convergence. This step is shown as follows:
\begin{equation}
    Z^{(n+1)} = \overline{U}^{(n+1)} + \frac{t^{(n)} - 1}{t^{(n+1)}} \big( \overline{U}^{(n+1)}- \overline{U}^{(n)} \big)
\end{equation} 
where
\begin{equation}
    t^{(n+1)} = \frac{1+\sqrt{1 + 4 \, (t^{(n)})^2}}{2},
\end{equation}
and $t^{(1)} = 1$. The updated $Z^{(n+1)}$ is then used in the next iteration to compute gradient step (\ref{eq:grad}) and the process is repeated until convergence. 
% Thanks to the above steps, FISTA emerges as an accelerated optimization algorithm encompassing forward-backward like iterations.
\begin{figure}
\vspace{-0.75cm}
    \centering
    \includegraphics[width=0.9\textwidth]{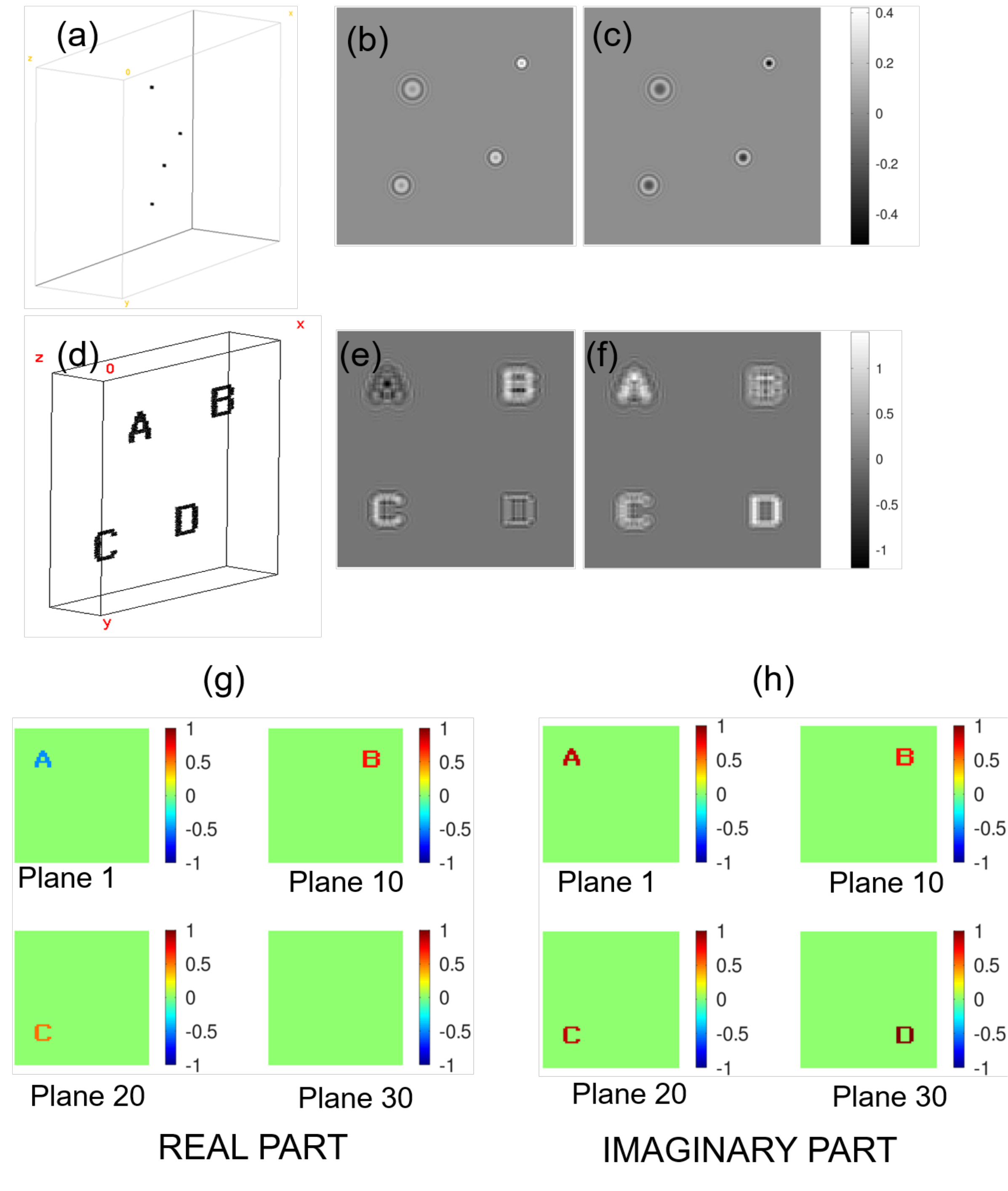}
    \caption{(a), (d): Magnitudes of amplitude and phase objects shown as 3D rendering, (b), (c): real and imaginary parts of the object field $V(x,y)$ at the detector plane due to the amplitude object, (e), (f): real and imaginary parts of the object field $V(x,y)$ at the detector plane due to the phase object, (g), (h): real and imaginary parts of the phase object for four specific planes where the letters A, B, C, D are located.}
    \label{fig:object_data}
\end{figure}
\section{Simulations and Results}
In simulations, we have considered a 3D computational box containing $128\times128\times30$ voxels which has equivalent physical dimension of $640\mu m \times 640 \mu m \times 750 \mu m$, and the illumination wavelength was assumed to be $0.5 \mu m$. The distance between the detector and the last plane of the object is set to be $1060 \mu m$. For the illustration, we take two 3D objects: the first is an amplitude object consisting of four small square reflectors located at four different planes (3, 11, 20, 28) of the volume, as shown in Fig.~\ref{fig:object_data}(a). Each object on a particular z-plane occupies a $2 \times 2$ pixel region. The second object is a phase object with four letters A, B, C, D placed in four different planes (1, 10, 20, 30), as shown in Fig.~\ref{fig:object_data}(d). The phase values for the four objects A, B, C, D are $2\pi/3$, $\pi/4$, $\pi/3$ and $\pi/2$ respectively and they all have unit amplitude. The real and imaginary parts of the object field $V(x,y)$ corresponding to the two objects as obtained by application of the operator $\hat{A}$ (Eq. (\ref{eq:A})) are shown in Fig.~\ref{fig:object_data}(b-c) and (e-f) respectively. In Fig.~\ref{fig:object_data}(g-h) we show the real and imaginary parts of the text phase object in four specific planes where the letters A, B, C, D are located. In a laboratory experiment, the object field $V(x,y)$ is not directly measured, but as explained before it is recoverable from an interference pattern $H(x,y) = |R(x,y)+V(x,y)|^2$ where $R(x,y)$ denotes the reference beam used for recording the hologram $H(x,y)$. Noise in the recovery of $V(x,y)$ is an important question. In a laboratory setting, noise in $V(x,y)$ will depend on the noise in the measurement of $H(x,y)$. As per nominal shot noise considerations, if sufficient light level is available for the recording of hologram $H(x,y)$ then highly accurate complex-field recovery is possible \cite{rajora2019mean}. The data incompleteness due to smaller number of detector pixels (compared to the number of voxels in the object volume) is therefore a much more important issue in this problem. As a result, we have not added any noise to $V(x,y)$.

The nature of the phase object demands some more discussion. A reflective phase object may be visualized if we assume the letters A, B, C, D to be mirrors coated with thin films of different thickness/refractive index. Another more realistic case for phase object is a transparent object like a biological cell. In such a case it will be more appropriate to have a transmission geometry. For a plane wave illumination, the effect of this on the data is that the object field $V(x,y)$ will have an additional constant background component \cite{Carter1970}. In the present work we assume that the background field at the detector plane has been removed from the recovered object field. This may for example be achieved by recording a blank hologram without any object, estimating the object field as a calibration step and then removing this background from the recovered object field $V(x,y)$ when the phase object is present. With this assumption the model represented by operator $\hat{A}$ in the previous section will still hold.   
Given the object field $V(x,y)$ at the detector plane, we solve the optimization problem (\ref{eq:cost_fun}) as explained in Section 3 to get an estimate $\overline{U}$ of the true 3D object. As previously mentioned, for the amplitude object in this illustration, we used $\ell_1$ sparsity prior with reality and positivity constraints. The regularization parameter $\alpha$ is chosen empirically to be $5\times10^{-4}$ for the amplitude object case.  
\begin{figure}
\vspace{-0.75cm}
    \centering
    \includegraphics[width=0.8\textwidth]{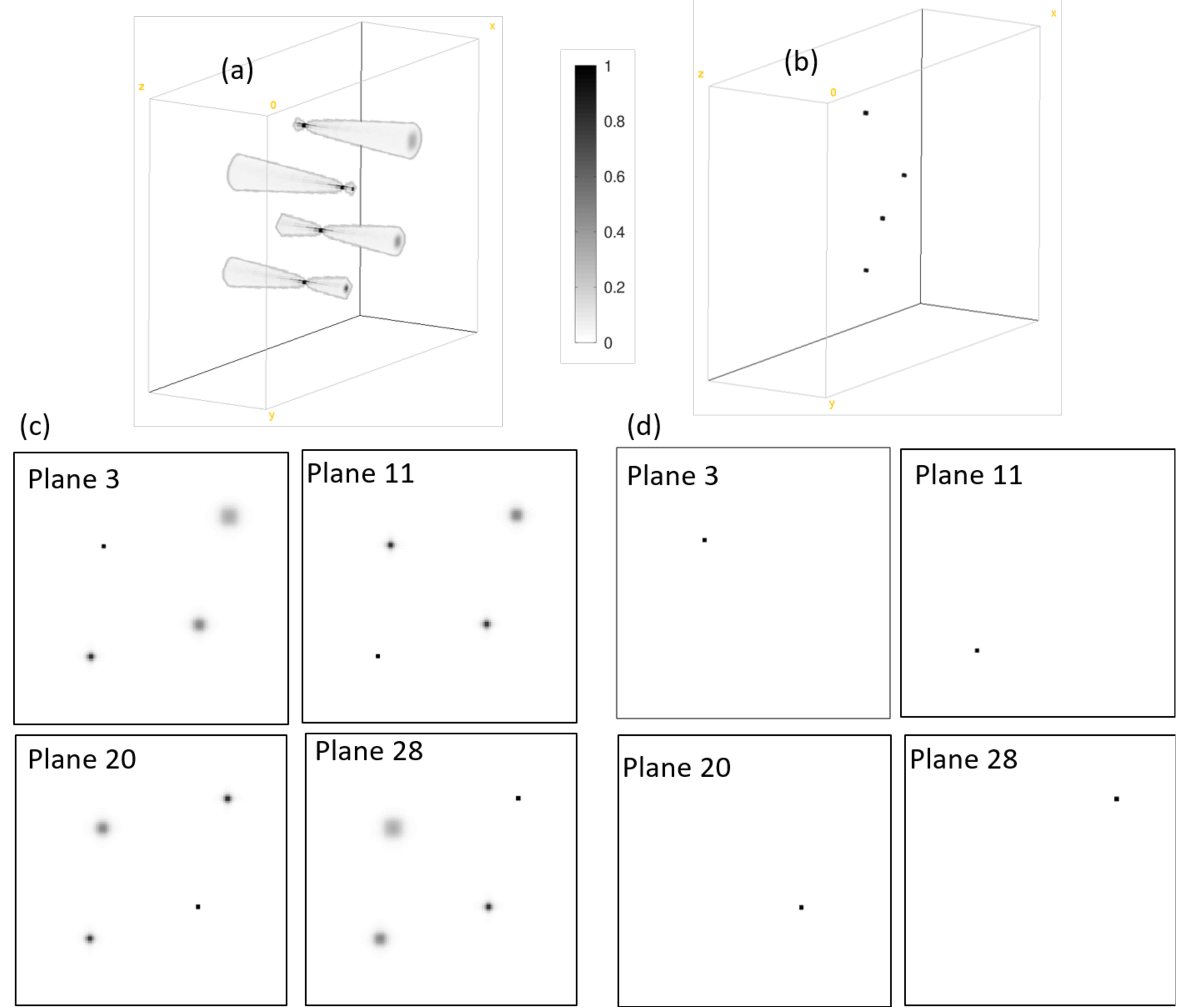}
    \caption{3D reconstruction of the amplitude object with four small reflectors. (a), (b): 3D rendering of the amplitude of the recovered four reflector object obtained by simple back-projection of the object field and by using the iterative reconstruction respectively. (c), (d): Image of the planes where the point objects originally existed in the 3D object box shown in (a) and (b) respectively.}
    \label{fig:amplitude_result}
\end{figure}
\begin{figure}
\vspace{-0.75cm}
    \centering
    \includegraphics[width=0.9\textwidth]{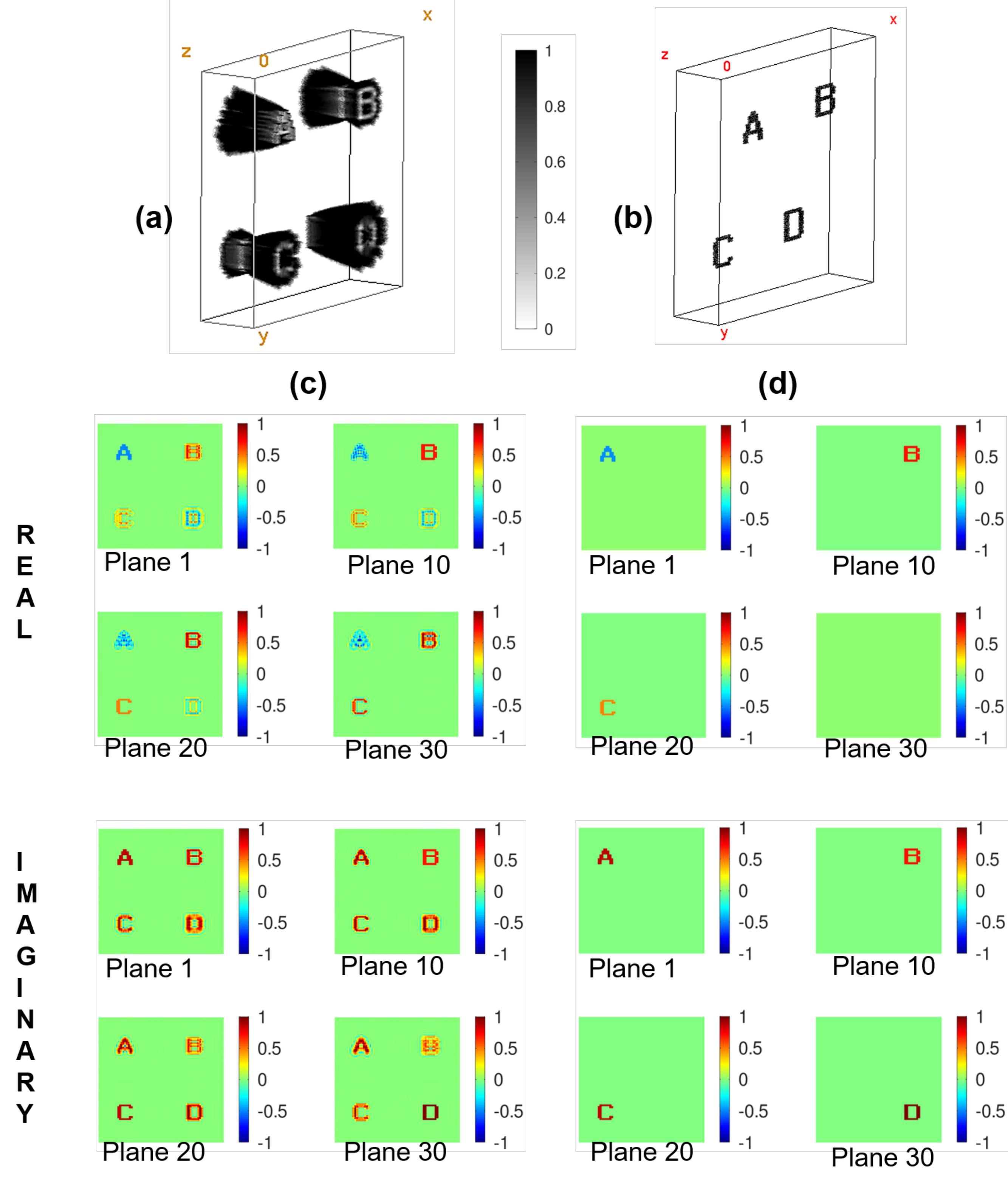}
    \caption{3D reconstruction of the text phase object. (a),(b): 3D rendering of amplitude of the reconstruction using simple back-projection and using proposed iterative reconstruction.  (c),(d): Real and imaginary parts for four different planes where the letters were located in the true 3D object box are shown for the reconstructions in (a), (b) respectively.}
    \label{fig:text_object}
\end{figure}
\begin{figure}
\vspace{-0.75cm}
    \centering
    \includegraphics[width=0.9\textwidth]{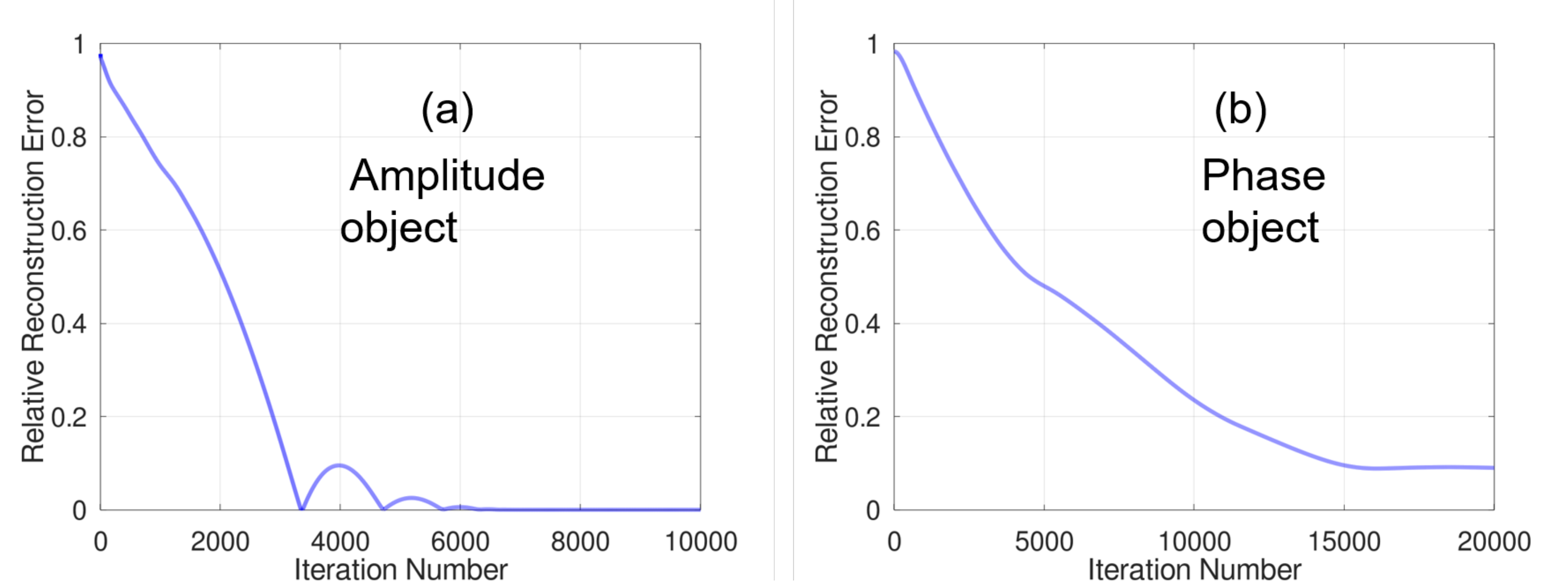}
    \caption{(a), (b): Relative reconstruction error for the two illustrations shown in Fig.~\ref{fig:amplitude_result} and Fig.~\ref{fig:text_object} as a function of iteration number respectively.}
    \label{fig:Error_plot}
\end{figure}
The corresponding amplitude recovery results are shown in Fig.~\ref{fig:amplitude_result}(b),(d). For comparison, we have also shown the amplitude object recovery by back-propagation of object field $V(x,y)$ to the different planes of the original 3D volume in Fig.~\ref{fig:amplitude_result}(a),(c). Similarly for the phase object, we show 3D rendering of the recovered object's amplitude and the real and imaginary parts of the recovery in the planes where the original letters A, B, C, D were located in Fig.~\ref{fig:text_object}(b),(d). This illustration uses the numerical value of regularization parameter $\alpha=10^{-3}$. Once again for comparison, the same quantities for simple back-projection of the object field are also shown in Fig.~\ref{fig:text_object}(a),(c). The 3D renderings shown in all the illustrations were generated using the Volume Viewer plug-in in the open-source software ImageJ. On visual inspection it may be observed that object recovery using the proposed iterative formalism closely matches with the ground truth 3D objects for both the illustrations. The simple back-projected fields over the original object volume however show non-zero de-focused fields in places where no object originally existed. In a MATLAB/Octave implementation of the code, for amplitude object we used $10,000$ iterations while for the text phase object we used $20,000$ iterations of the proposed algorithm. 
 
The relative error in the object domain with respect to the ground truth objects, namely the reconstruction error, is given by
\begin{equation}
    E^{\rm{(object \;domain)}} = \frac{|| U_{\rm{(ground \;truth)}} - \overline{U}||_2}{|| U_{\rm{(ground \;truth)}} ||_2},
\end{equation}
and is plotted for the two cases in Fig.~\ref{fig:Error_plot}(a),(b) as a function of iteration number. The relative error for the amplitude object was down to approximately $0.016$ but for the text phase object was seen to be close to $0.095$. The main reason for this is that the $\ell_1$-norm penalty amounts to soft-thresholding which makes a large number of voxels in the reconstructed volume go identically to zero. For the case of the text phase object, only TV regularization was used. Careful examination of the solution volume as in Fig.~\ref{fig:text_object} showed that the voxels in the iterative solution where no object originally existed did not identically become zero but had a small numerical value that was three orders of magnitude lower than the magnitude of the text voxels. It may be possible to use $\ell_1 + TV$ penalty in the second case as well to make these voxels go to zero identically thereby improving the relative error performance. The relative object domain errors for the simple back-projection based reconstruction are however quite large and equal to $4.32$ and $5.39$ for the amplitude and phase objects respectively. For both the illustrations, the data domain error
\begin{equation}
    E^{\rm{(data \;domain)}} = \frac{|| V - \hat{A} \overline{U}||_F}{|| V ||_F},
\end{equation}
was of the order of $10^{-4}$ at the end of the iterations. The 3D reconstructions using a single object field data frame in the hologram plane is possible as shown in these illustrations
since both the 3D objects were sparse. True 3D reconstructions may not be possible for non-sparse objects from degrees-of-freedom considerations. In such cases, additional diversity mechanisms  may be required for measuring non-redundant data. The question of whether true 3D reconstruction is always possible is however not just mathematical in nature, as the assumptions about small secondary scattering that have been used here are not valid as the 3D object gets more and more complex in its structure. These considerations need substantial further study before making any quantitative statement regarding possibility of reconstructing a 3D object of given complexity. Nevertheless the iterative approach presented here may be sufficient for 3D reconstruction of low complexity objects like transparent cells, optical elements like lenses, etc. We will report on experiments using the proposed methodology for such objects in future.

\section{Conclusion and Discussion}
In conclusion, we have examined the problem of 3D image reconstruction in holography. The traditional holographic replay was identified to be the Hermitian transpose operation corresponding to the forward hologram formation process. Using this knowledge, we constructed an iterative solution to the 3D reconstruction problem which used object sparsity as a constraint. The resultant 3D reconstruction is seen to be a significantly better representation of the original 3D object as compared to the simple back-projection based image reconstruction. We note that the 3D reconstruction using the constrained optimization strategy as proposed here cannot be achieved by any form of physical holographic replay. The methodology therefore truly takes full advantage of the fact that complex object wave at the detector plane is available in numerical form in digital holography setups.  The technique may find important applications in a number of digital holographic or interferometric imaging systems as we will report in future.

\bibliographystyle{unsrt}  
\bibliography{References}  

\end{document}